\documentclass[pra,superscriptaddress,reprint,twocolumn]{revtex4-2}
\usepackage{amstext}
\usepackage{amssymb} 
\usepackage{amsmath}
\usepackage{amsthm} 
\usepackage{algorithm}
\usepackage{algpseudocode}
\usepackage{hhline}
\usepackage{dsfont}
\usepackage{float}
\usepackage{tikz}
\usepackage{makecell}
\usepackage{siunitx}
\usepackage{bbold}
\usepackage[braket,qm]{qcircuit}
\usepackage{braket}
\usepackage{mathtools} 
\usepackage{xcolor}
\usepackage{hyperref}
\usepackage{enumitem}
\usepackage{natbib}
\hypersetup{colorlinks=true,allcolors=[rgb]{0.1,0.1,0.5}}

\begin{document}

\title{Physical limits of ultra-high-finesse optical cavities: Taming two-level systems of glassy metal oxides}
	
\author{H. R. Kong}
\affiliation{Institute for Quantum Computing and Department of Physics \& Astronomy, University of Waterloo, Waterloo, Ontario N2L 3G1, Canada}
\affiliation{Q-Block Computing Inc., Kitchener, Ontario N2C 2C8, Canada}
\author{K. S. Choi}
\email{kyung.choi@uwaterloo.ca}
\affiliation{Institute for Quantum Computing and Department of Physics \& Astronomy, University of Waterloo, Waterloo, Ontario N2L 3G1, Canada}
\affiliation{Q-Block Computing Inc., Kitchener, Ontario N2C 2C8, Canada}

\begin{abstract}
The framework of tunnelling states in amorphous media provides the dissipative mechanism that imposes fundamental limits on the performances of ultra-high-Q optical and microwave resonators. In the optical domain, however, the microscopic nature of the tunnelling states and their direct consequences to the optical losses have not been characterized. In this work, we examine the interplay between glassy physics and stoichiometric point defects of ultra-low-loss dielectric mirrors, with a focus on two-level systems of oxygen defects in metal oxide, Ta$_2$O$_5$. In the context of ultra-high-finesse optical cavities, we characterize the Urbach tail of Ta$_2$O$_5$ under thermal annealing and reactive oxygen plasma. For the first time, we microscopically resolve the individual oxygen point defects on the mirror surfaces and introduce laser-assisted annealing to selectively ``cure" surface defects. By exploiting these techniques, we report the realization of ultra-high-finesse optical cavity with finesse $\mathcal{F}=450,000$ and the absorption-scatter loss $ 0.1 \pm 0.2$ ppm at $852$ nm.
\end{abstract}

\maketitle
\section{Introduction} 

While mechanical properties of crystalline structure are well understood, amorphous media have more complex properties. Due to the extensive metastable states, glassy systems are prevented from equilibration to the thermal bath, and their structural organizations manifest bulk properties that can be measured by specific heat and thermal conductivity. The dynamical behaviour of amorphous (glassy) systems at low-temperature can be captured by the phenomenology of low-energy two-level systems (TLS) \cite{Phillips1972, Anderson1972,Caldeira1983,Phillips1987} that fluctuate and dissipate the spatial ordering of the glass. In this context, annealing could be thought of as a method to equilibrate their configuration states and to relax towards lower energies. 

\begin{figure*}[t!] 
\centering
\includegraphics[width=1.5\columnwidth]{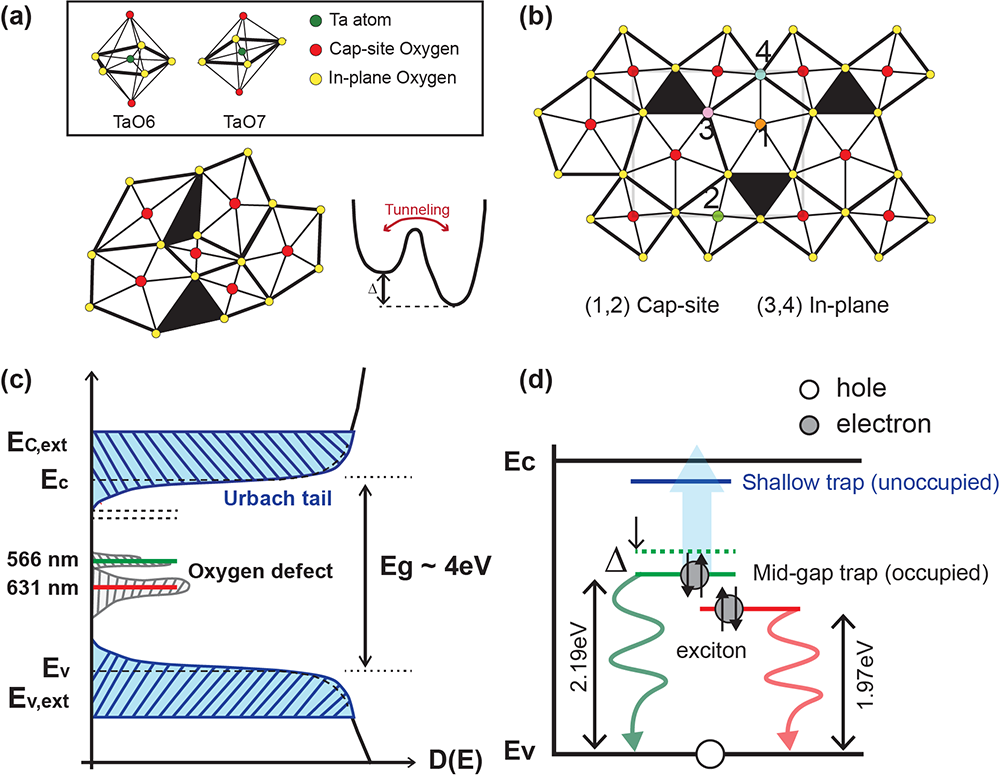} 
\caption{Two-level systems and point defects in amorphous (a-Ta$_2$O$_5$) and crystalline tantalum pentoxide ($L$-Ta$_2$O$_5$). (a) a-Ta$_2$O$_5$. Long-range structural disorder can occur between configuration states. Such structural organization is phenomenologically mapped to a model of a thermally-activated tunnelling between two coupled potential wells. (b) $L$-Ta$_2$O$_5$ \cite{Stephenson1971,Yang2018}. (inset) Building block of a-Ta$_2$O$_5$ and $L$-Ta$_2$O$_5$. The relaxed crystal structure consists of several building blocks in corner (edge) sharing, with the Ta located at the center of the polyhedra. The inset displays the two polyhedras: Octohedra TaO$_6$ (left), and pentagonal bipyramid TaO$_7$ (right). Green sphere represents the tantalum atom, while the yellow (red) spheres are the oxygen atoms on the in-plane (cap) sites. The four dominant locations of oxygen vacancies are labeled with their indices and color coded. (c) Density of states (DOS) for a-Ta$_2$O$_5$. The Urbach tails (broadened oxygen defect states) are shown due to their long-wavelength (localized) structural disorder. $E_c$ ($E_v$) is energy for the conduction (valence) band. The extended states of the conduction (valence) band formed by the long-range structural organizations are denoted as $E_{c,ext}$ ($E_{v,ext}$). (d) Possible optical transitions in in-plane oxygen vacancies in a-Ta$_2$O$_5$. The charge state is occupied in the mid-gap trap, and can be excited to the conduction band to generate photocurrent or to the shallow trap, which subsequently decays to the mid-gap trap. In photoluminescence (PL) spectroscopy, exciton generated near $2.19$ eV by $450$-nm pump laser experiences electron-hole pair recombination, generating PL photons at $566$ nm ($2.19$ eV) and $631$ nm ($1.97$ eV).} 
\label{fig1} 
\end{figure*} 

Since the $70$s, the semiconductor industry and related areas have made a broad scope of studies around dielectric materials, with a focus of its dielectric loss that relates to the performance of field effect transistors \cite{Sawada1999,Kim2013}. In the modern context, TLS of amorphous metal oxides is the limiting factor for the performances of gravitational wave detector and optical clocks by their coating brownian motion \cite{Martin2014,Cole2013,Amato2018,Amato2019, Amato2020}, and can also contribute to the optical losses for cavity quantum electrodynamics (QED) experiments \cite{Rempe1992}. The TLS in native oxide layers of active metals causes an increased micromotion for the trapped ions on surface traps \cite{Brownnutt2015} and limits the $Q$-factor of superconducting circuits with junction noise \cite{Muller2019}. For ultra-high-$Q$ optical resonators, structural organization of high-index metal oxides gives rise to bulk optical properties near the mobility edge from the extensive localized states that deform their optical density of states with a finite Urbach tail. In addition to the long-range order, interfacial local defects can dissipate and fluctuate the optical fields in high-finesse cavities over a spectral extent connected to their short-range ordering. The structural disorders pose a fundamental limit to the optical $Q$-factors of optical cavities and interferometers that can be attained by amorphous media. As a premier material for ultra-low-loss optical interferometers over the past decades, most studies in the optical domain have exclusively focused on the long-wavelength structural disorders of tantalum pentoxide (Ta$_2$O$_5$, hereafter referred as tantala) that give rise to the coating thermal noise of ultra-high-finesse optical resonators \cite{Harry2006,Amato2018,Amato2019, Amato2020, Kessler2012, Andrew2015}. However, the microscopic origins of the TLS and their direct manifestations to the optical losses are closely connected to the short-wavelength structural organization, and have not yet been characterized in any experiment.

In this work, we investigate the crucial role of amorphocity on the dielectric coatings of optical mirrors with a focus on tantala deposited by ion-beam sputtering (IBS). We perform ellipsometric measurements of the Urbach tails of ultra-low-loss optical films and investigate the thermal annealing and O$_2$ plasma treatments for the reduction of optical loss in the UV range. We also spatially resolve the individual oxygen point defects at the interfaces of bilayer quarter-wave stack for ultra-low-loss mirror with a high-resolution optical microscope, and confirm their defect states near the Fermi level $2.19$ eV by photoluminescence emission spectroscopy. Importantly, we examine the oxygen vacancy as an optically active TLS that limits the optical $Q$-factor of ultra-high-finesse cavities in the near-infra-red (NIR) regime. We develop a microscopic technique of laser-assisted local annealing for the removal of the stoichiometric defects at the atomistic level. By constructing ultra-high-finesse optical resonators, we observe the key role of oxygen vacancy to the mirror's absorption-scatter ($AS$) losses. We thereby report the realization of a nearly defect-free ultra-high-finesse optical cavity for the recent many-body QED experiment \cite{Kong2021} with optical finesse $\mathcal{F}=445,000$ and $AS=0.1\pm 0.2$ ppm at $852$ nm. Our result improves significantly upon the seminal work by Rempe \textit{et al.} \cite{Rempe1992}, which have held the lowest AS $\sim 1.1$ ppm for any optical device ever since 1991.

\section{Noise model of amorphous tantala}\label{noise_section}

Unlike crystalline structures with well-defined valence and conduction bands, amorphous dielectrics are characterized by the absence of long-range ordering (LRO). However, they have short-range ordering (SRO) at the molecular level, which opens a gap similar to crystalline materials. A prominent example of TLS defects in amorphous oxides is attributed to hydrogen in interstitial positions \cite{Gordon2014}. While they bond to the primary oxygen, they can also migrate to the neighbouring atoms. Following the TLS convention \cite{Phillips1972, Anderson1972,Caldeira1983,Phillips1987}, this process can be captured by thermally activated tunnelling in an asymmetric double-well potential, as shown for amorphous tantala (a-Ta$_2$O$_5$) in Fig.\ref{fig1}(a). As the hydrogen is the dominant background gas in ultra-high-vacuum environment, it is not always possible to prevent hydrogen contamination in thin film deposition and leads to an increased Brownian motion of the coatings. 

The dynamics of TLS causes an internal friction to the dielectric media, which can be detected by the phonon scattering relaxation \cite{Cahill1989,Liu1998,Classen2000,Pohl2002}. Namely, an internal friction is proportional to the tunnelling rate between the two potential well of the TLS modes. Mechanical loss of various metal oxide dielectrics have been extensively surveyed \cite{Pohl2002} and high-sample temperature and Ar-bombardment have been shown to be effective against oxygen deficiency that leads to internal friction in reactive IBS deposition \cite{Liu2014}. While the microscopic origins of the TLS remain unclear, the compositional disorder by oxygen vacancies is a possible source of TLS for the mechanical loss in a-Ta$_2$O$_5$. Such a thermal noise of high-index tantala thin film ($n_H=2.1$ at $852$ nm) limits the sensitivity of present gravitational wave detectors \cite{Harry2006} and the stability of optical references in an optical clocks \cite{Kessler2012,Andrew2015}. In the context of a-Ta$_2$O$_5$, mechanical dissipations can be alleviated by Titania-doping (Ti:Ta$_2$O$_5$) as network modifier \cite{Manchanda2001,Atanassova2010} and thermal annealing \cite{Vajente2018,Amato2018}. Alternative high-index ($>n_H$) metal oxides are being developed with the objective of reducing the thicknesses of the dissipative coating layers \cite{Amato2020}. Recent efforts include the development of crystalline mirrors to eliminate the structural TLS defects from the coating \cite{Cole2013}. 

Beyond the thermal noise, the structural disorder introduced by oxygen defects in the LRO also contributes to optical losses of amorphous tantala \cite{Markosyan2013}. Indeed, ab-initio studies confirm the microscopic origins of Urbach tails to structural topological organization caused by the relaxation in the presence of disorder \cite{Pan2008}. As illustrated in Fig. \ref{fig1}(c), near the conduction band edge, the extended states become localized and their optical density of states $DOS(E)$ are broadened by an exponentially decaying Urbach tail $DOS(E)\propto e^{-(E_{gt}-E)/E_{U}}$, where $E_U$ is the Urbach energy and $E_{gt}<E_{c,ext}$ characterizes the beginning of the Urbach tail for the extended states $E_{c,ext}$. The Urbach tail accompanies an optical absorption that decays from the extended states over the energy scale $E_U$, caused by the structural and thermal disorder in the system \cite{Pan2008}. Such disorder creates a sub-gap broadening of the absorption edge that leads to an optical loss within $E_U$ and detected by spectroscopic ellipsometry. With the band gap $\sim 4.2$ eV ($302$ nm) of tantala, the finite Urbach tail can be the dominant source of the optical absorption observed in high-finesse cavities in the UV and visible wavelengths $<450$ nm. 

Despite the amorphous nature of tantala, they also exhibit SRO at the molecular level, and the oxygen vacancies constitute microscopic point defects of the polycrystal. Similar to defect states in optical crystals (e.g., nitrogen vacancies in diamond), oxygen vacancies generate well-defined bound states (trap levels) and permit direct optical transitions. To understand the physics of these transitions and their intrinsic optical losses, let us briefly review the molecular structure of tantala and its oxygen defects. In tantala, there are $4$ types of oxygen vacancies shown in Fig. \ref{fig1}(b) that depend on their locations. As shown in Fig. \ref{fig1}(b), the oxygen vacancies can arise from the absence of in-plane oxygen or cap oxygen, and they display qualitatively different optical behaviour from each other \cite{Pan2008}. Since the cap oxygen vacancies are only shared by two polyhedra, the defect formation energy $E_d^{\text{cap}}\sim 4.9-5.3$ eV is considerably higher than in-plane oxygen vacancies $E_d^{\text{in-plane}}\sim 3.8-4.3$ eV \cite{Yang2013}, and oxygen defects are thereby more likely to relax into in-plane vacancies. Indeed, cap oxygen deficiencies (labeled as type $1$ and $2$ in Fig. \ref{fig1}(b)) create trap levels close to and buried within the conduction band, and are not relevant to our study. In-plane vacancies, such as type $3$ ($4$), is a deficiency of a single neutral oxygen that was originally shared with one (two) octohedra TaO$_6$ and two (one) pentagonal bipyramid TaO$_7$. Vacancies thereby leave behind the dangling bonds at the defect sites for the two electrons and form a well-defined two-level system (mid-gap trap and shallow trap) near the Fermi level hybridized by Ta (5d orbital) and O (2p orbital) within the bandgap of the tantala, which is optically addressable. Time-dependent density functional theory indicates that the occupied trap levels of type $3$ and $4$ defects are approximately located at $2.19$ eV ($566$ nm) and $1.97$ eV ($631$ nm) above the valence band $E_v$ for dilute (non-interacting) defects \cite{Ramprasad2003}, as shown in Fig. \ref{fig1}(d). 

\section{Long-range structural organization}\label{ellipsometer_section}

While our primary focus is the reduction of optical loss, extensive research at LIGO \cite{Amato2018,Amato2019,Amato2020} confirms the correlation between Urbach tail and internal friction in the metal oxides. Moreover, beyond its role as a network modifier, the titania doping is thought to quench the danging bonds of excess oxygen. Indeed, oxygen vacancies of tantala can be compensated by Ti$^{4+}$ ions as they substitute into the Ta$^{5+}$ sites. Thermal annealing in oxygen rich environment has shown effective in the structural ordering of tantala. Reduction of unpaired electrons from the oxygen vacancies has been detected by a $^{17}$O MAS NMR spectra post thermal annealing \cite{Kim2011}. We thereby apply a two-step post-treatment for the IBS mirrors: (1) Thermal annealing to relax the structural organization, followed by (2) a reactive O$_2$ plasma to reduce stoichiometric oxygen defects.

Bilayer quarter-wave stack  ($34$-layers) of tantalum pentoxide and silica was deposited on a superpolished fused silica substrate with RMS surface roughness $\sim 0.25 \mathring{A}$  by a reactive IBS with Ar$^+$ bombardment. At the design wavelength $852$ nm, the coating thickness for tantala (silica) is $162$ ($268$) nm. The IBS mirrors were packaged in the cleanroom and annealed within a fused silica container. Based on Refs. \cite{Kim2013,Amato2019,Amato2020}, we perform an atmospheric thermal annealing at target temperature $450$ $^{\circ}$C in a muffle furnace, well below the critical temperature $T\sim 650 ^{\circ}$C for the crystalline phase of tantala that yields an increased optical scattering loss by the grain boundaries \cite{Vajente2018,Amato2018}. To relieve the coating stress, the ramping schedule was limited to 20 $^{\circ}$C/hr and the sample was held at the target temperature for $12$ hour. Following the annealing, the sample was treated by a reactive oxygen plasma ($50$ W at $50$ kHz) at base pressure $10$ mTorr and O$_2$ flow rate $5$ scm.

\begin{figure*}[t!]
\centering
\includegraphics[width=1.5\columnwidth]{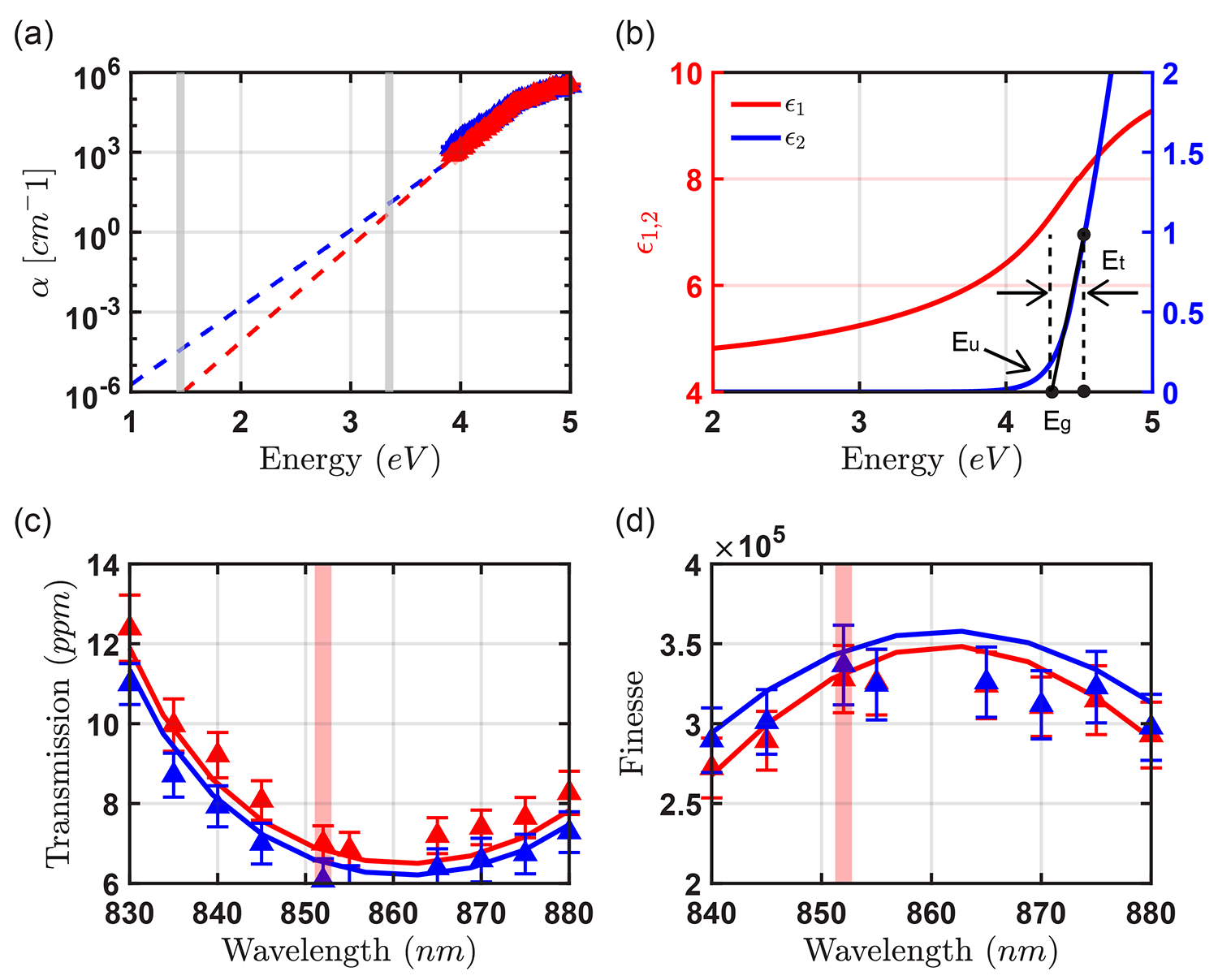}
\caption{Relaxing structural organizations by thermal and plasma annealing. (a) Measured absorption coefficient $\alpha$ of Ta$_2$O$_5$ from the ellipsometric data between $3.9$ eV and $4.3$ eV. The fitted curves are based on the Cody-Lorentz formula. (b)  Reconstructed complex permittivities $\tilde{\epsilon} = \epsilon_{1} + i \epsilon_{2}$ of tantala. Only the data points for the complex permittivity before the annealing is shown. (c) Measured mirror transmission ($T$) based on loss partition in cavity ring-down spectroscopy \cite{Hood2001}. The curves are computed from the transfer matrices with the ellipsometric determination of $\tilde{n}$ for the coating layers. (d) Optical finesse of the Fabry-Perot cavity. The finesse is determined from the total loss $\mathcal{L}=T+AS$, where $AS$ is the absorption-scatter loss. For the theory curve, in addition to the transfer matrices from the ellipsometric data, we use the absorption-scatter loss as a fitting parameter. Blue (red) data is measured before (after) the annealing. The annealing process consists of thermal annealing at atmospheric pressure, followed by a reactive O$_2$ plasma treatment. Urbach energy before (after) annealing is $150$ meV ($121$) meV, and the gap energy before (after) annealing remains the same at $4.2$ eV.}
\label{fig2}
\end{figure*}

\begin{figure*}[t!]
\centering
\includegraphics[width=1.5\columnwidth]{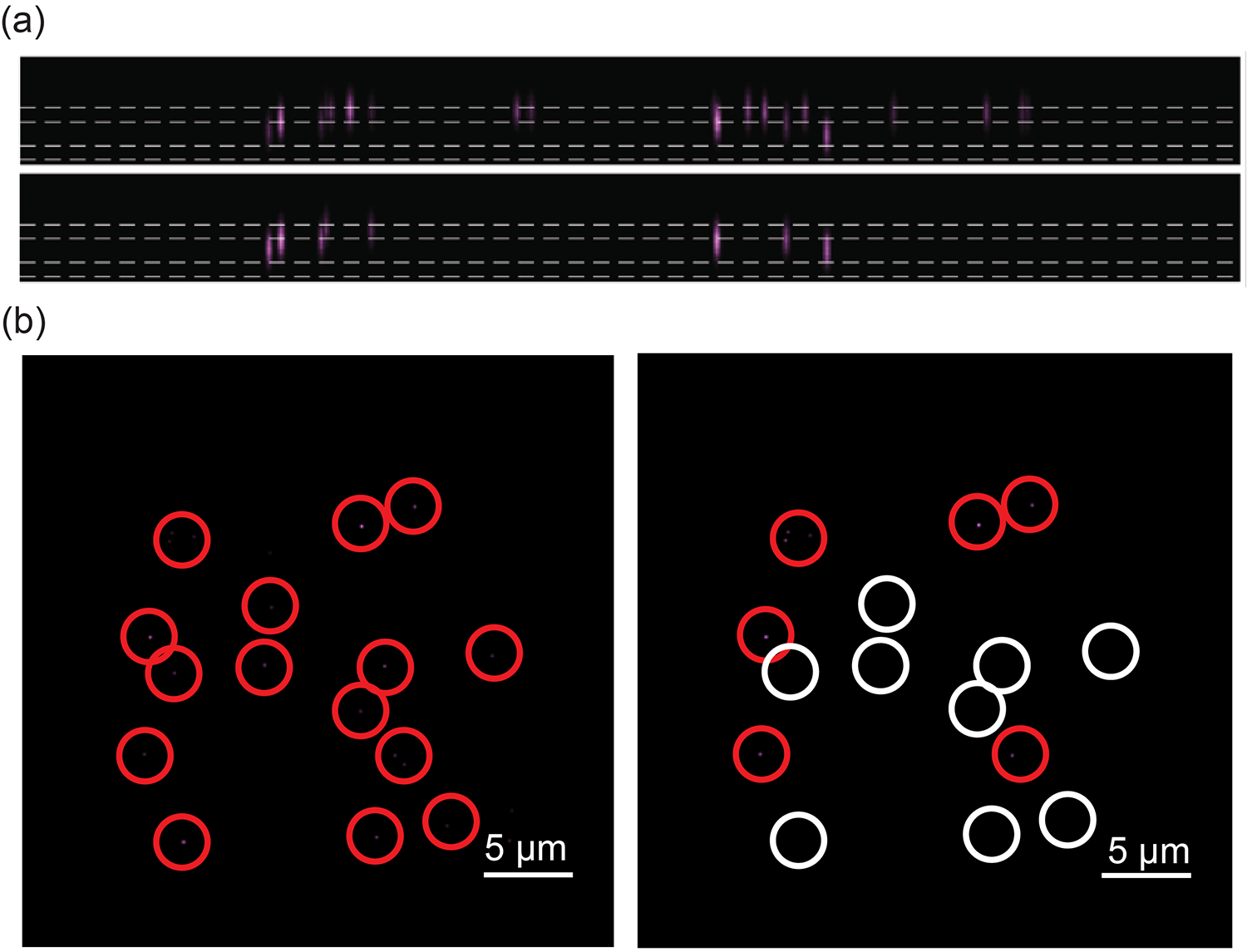}
\caption{Superresolution imaging of individual oxygen vacancies and laser-assisted annealing. (a) Cross-section of the reconstructed 3D image (depth image). The depth image before (after) laser-assisted annealing is shown on the top (bottom). The oxygen defects are mostly interfacial between tantala and silica (or vacuum). (b) Cross-section of reconstructed 3D image (transverse image). The individual oxygen vacancies are labeled with red circle before (after) the annealing on the left (right). The white circle indicates the position of oxygen defects that have been annealed. By inducing photoluminescence emission, 3D image of the individual defects is reconstructed from deconvolution microscopy that localizes the emitters to transverse ($10$ nm) and depth ($75$ nm) resolutions \cite{He1997}. The point spread function is determined by a fluorescence bead on a test target. }
\label{fig3}
\end{figure*}

To characterize the Urbach tail, we perform a spectroscopic ellipsometry across a broad spectral range for the tantala \cite{Urbach1953,Fujiwara2007,Cody1981}. We independently characterize the coating thicknesses from transmission measurements with a laser-based spectrophotometer and measure the complex refractive index of IBS-grown silica as a reference. In particular, we detect the reflected fields $r_p,r_s$ of the sample at incident angle $55^{\circ}$ and reconstruct the ellipsometric angles $(\Psi,\Delta)$ where $\rho=r_p/r_s=\tan(\Psi)\exp(i\Delta)$ and $r_p$ ($r_s$) is the reflected field projected to the $p$ ($s$) polarization basis. From ($\Psi, \Delta$), we can reconstruct the refractive index $\tilde{n}_H= n_H+i k_H$ of high-index dielectric layers from the optical transfer matrices, assuming known coating thickness. We thereby obtain the Urbach energy $E_U$ by fitting the complex permittivity $\tilde{\epsilon} = \epsilon_{1} + i \epsilon_{2}= \tilde{n}^2$ using a Kramers–Kronig compliant Cody-Lorentz (CL) equation.

The CL model integrates the Lorentz oscillator with optical absorption near the extended states due to their structural organizations (such as the Urbach tail) and is given by
\begin{equation}
\begin{split}
&\epsilon_{1}(\omega)-1= \frac{2}{\pi} \oint  \frac{x \epsilon_{2}(x)}{x^2- \omega^2} dx\\ 
&\epsilon_{2}= \left\{ 
\begin{array}{l} 
\frac{E_{\text{gt}} G(E_{\text{g}}) L(E_{\text{g}})}{E} ~ \text{exp} \left( \frac{E-E_{\text{gt}}}{E_{\text{u}}} \right), 0< E \leq E_{\text{gt}}    \\
G(E) L(E) , ~~~ E > E_{\text{gt}}
\end{array} 
\right.
\label{eqn:epsilon2CL}
\end{split}
\end{equation}
where $E_{\text{gt}}=(E_{\text{g}}+E_{\text{t}})$ and gap energy $E_g$. For $E > E_{gt}$, $\epsilon_{2}$ is a product of a variable band edge function $G(E)=\frac{(E-E_{\text{g}})^2}{(E-E_{\text{g}})^2+E_{\text{P}}^2} $ and Lorentz oscillator $L(E)=\frac{A E_{0} \Gamma E}{(E^2-E_{0}^2)^2+\Gamma^2 E^2}$. For Lorentz oscillator function, $E_{0}$ is the peak energy for the Lorentz oscillator, $A$ is the amplitude of the Lorentz oscillator, and $\Gamma$ is the width of the Lorentz oscillator. The transition energy $E_{P}$ defines the weighting factor between the Cody and the Lorentz models. 

Fig. \ref{fig2}(a)-(b) show the spectroscopic ellipsometry measurements. Blue (red) data points are the measurement results before (after) the annealing. Before the annealing, we observe gap energy $E_{\text{g}} \simeq 4.2\pm 0.025$ eV and Urbach energy $E_U^{\text{initial}}=150\pm 17$ meV, in correspondence to other stoichiometric IBS tantala \cite{Amato2018,Amato2019,Amato2020}. $E_g^{\text{initial}}$ and $E_U^{\text{initial}}$ are marked in the complex permittivities of Fig. \ref{fig2}(b). After the thermal annealing, we observe the structural relaxation accompanied by the decrease in the Urbach energy $E_U^{\text{anneal}}=121\pm 10$ meV. Fig. \ref{fig2}(b) displays the improved absorption coefficient $\alpha$ after annealing, extrapolated from the Urbach tail of the fitted CL model for the ellipsometric data between $3.9-4.3$ eV. 

To characterize the optical loss around $852$ nm, we perform a loss partition of the cavity mirror by analyzing the transmitted and reflected intensities of the probe field during a cavity ring-down spectroscopy \cite{Hood2001} (Section \ref{UHFcavsection}). The finesse $\mathcal{F}=\pi/\mathcal{L}$ is computed based on the total loss $\mathcal{L}=T+AS$ of the mirror. We observe the decrease of absorption-scatter loss from $AS=3.2\pm0.5$ ppm to $AS=2.6\pm 0.4$ ppm. Due to the improved stoichiometry of tantala, the refractive index $n_H$ is reduced and results in higher transmission from $T=6.1\pm0.5$ ppm to $T=7.0\pm0.4$ ppm, leading to a minute reduction in the optical finesse (Fig. \ref{fig2}(d)). Other mirror samples consistently produce similar absorption-scatter loss $AS\sim 2-2.5$ ppm after the annealing, regardless of their initial $AS$. This seems to suggest that our annealing process brings the tantala coating close to a stable optimal configuration. 

\begin{figure*}[t!]
\centering
\includegraphics[width=1.5\columnwidth]{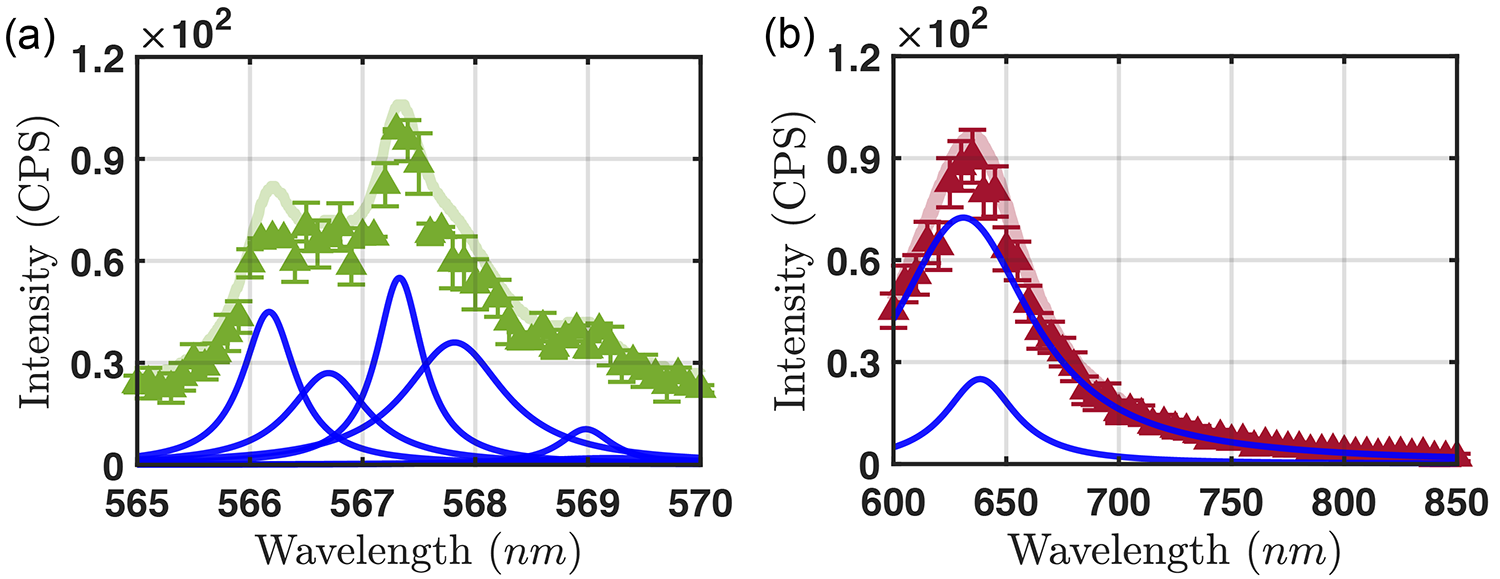}
\caption{Photoluminescence spectroscopy. Mid-gap trap levels are spectroscopically resolved. The spectrum is analyzed by the Voigt method and divided into multiple peaks with (a) center wavelengths near $567$ nm and (b) $640$ nm.}
\label{fig4}
\end{figure*}

A couple of remarks are in order. As a reference, we indicate two wavelengths for consideration in Fig. \ref{fig2}(a). The first is at $852$ nm for our many-body QED platform \cite{Kong2021}. In the NIR range, the reduced absorption from $\alpha\simeq 4.5 \times 10^{-5}$ cm$^{-1}$ to $< 10^{-6}$ cm$^{-1}$ does not have a meaningful contribution to the observed reduction of $AS$. Using the ellipsometric data, the expected absorption loss by the Urbach tail is only $AS\sim 0.02$ ppm for our mirror at $852$ nm before the annealing. From AFM scans, we estimate the scattering loss $S\sim 0.14$ ppm with RMS surface roughness $\sim 0.25\mathring{A}$. On the other hand, for the near-UV cavity QED \cite{Brandstatter2013,Gangloff2015}, we anticipate that the absorption loss caused by the Urbach tail could be as high as $AS=290$ ppm ($\alpha=34$ cm$^{-1}$) at $370$ nm and $AS=28$ ppm ($\alpha=1.9$ cm$^{-1}$) after annealing. These predictions are consistent to the absorption-scatter losses of high-finesse UV cavities in the literature \cite{Brandstatter2013,Gangloff2015}. Our spectroscopic study of the Urbach tails indicates the relevance of long-range structural organization for short-wavelength high-finesse optical cavities.

\section{Photoluminescence spectroscopy and superresolution defect imaging}
Far away from the extended states of the conduction and valence bands (such as in the NIR range), optical losses of amorphous dielectrics are dominated by direct dipole-allowed transitions from (to) localized states to (from) extended states near the band edges. In particular, we focus on the in-plane oxygen vacancy levels, which form mid-gap trap levels near the Fermi level (See Fig. \ref{fig1}(c)-(d)). Because of the short-range nature, the emission linewidth $E_d$ of photoluminescence is anticipated to be broad $E_d>E_U$, generating photons around $566$ nm ($2.19$ eV) and $631$ nm ($1.97$ eV). They may have sufficiently large dipole moment to dissipate the optical field in the cavity off-resonantly.

To develop a phenomenological picture, let us consider the TLS (radiative rate $\Gamma$) formed by the mid-gap trap $|g\rangle$ and the extended states $|e \rangle$ of the valence band with the TLS-cavity Hamiltonian
\begin{equation}
H_{\text{int}}= g_d (\hat{a}^{\dagger} \sigma_{\text{ge}}+\hat{a} \sigma_{\text{eg}}), 
\end{equation}
in the rotating frame of the cavity field's frequency, where $g_d$ is the coupling rate of the cavity field to the defect and the TLS Hamiltonian $H_{TLS}=\Delta_d|e\rangle\langle e|$. For large detuning $\Delta_d\gg g_d,\Gamma$, the intrinsic cavity loss $\kappa_{\text{int}}\simeq (g_d/\Delta_d)^2\Gamma$. We thereby anticipate the absorption loss of the cavity to slowly vary from the resonance with $\sim 1/\Delta_d^2$. Hence, in close spectral proximity to the defect states, the oxygen vacancies may have a decisive role in determining the absorption loss of ultra-high-finesse optical cavities at the few ppm level.

Indeed, there have been reports of mirror degradation by oxygen deficiencies under UHV environment \cite{Brandstatter2013,Gangloff2015,Cetina2013}. In particular, Ref. \cite{Gangloff2015} has claimed that the absorption-scatter loss $AS$ from oxygen vacancy may be the primary contributor to the observed increase of $AS$ when the optical cavity is exposed to UV field over a prolonged period. As discussed in Section \ref{noise_section}, the defect formation energies of in-plane oxygen vacancies greatly exceed the laser frequencies used in these experiments, and it is unlikely that oxygen defects could have been formed by the lasers. Indeed, recent work \cite{Schmitz2019} proposes the UV-enhanced deposition of hydrocarbons, generated from organic compounds, as the mechanism in which optical finesse in Ref. \cite{Gangloff2015} decays under vacuum.

To make progress, we utilize a high-resolution UV optical microscope with $NA=0.75$ and spatially resolve the individual oxygen defects on the IBS mirrors after the aforementioned annealing post-processing (Section \ref{ellipsometer_section}). We excite the mirror substrate with $450$ nm pump with laser power between $5-30$ mW to generate photoluminescence (PL) photons by exciton recombination near the mid-gap trap levels. We collect the PL photons on an EMCCD camera after a $550$ nm edge-pass filter. We generate $61$ images every $500$ nm steps and the 3D image stack was processed with deconvolution algorithm \cite{Wendykier2010} to reconstruct Fig. \ref{fig3} with transverse (longitudinal) resolution $10$ nm ($75$ nm).

\begin{figure*}[t!]
\centering
\includegraphics[width=1.75\columnwidth]{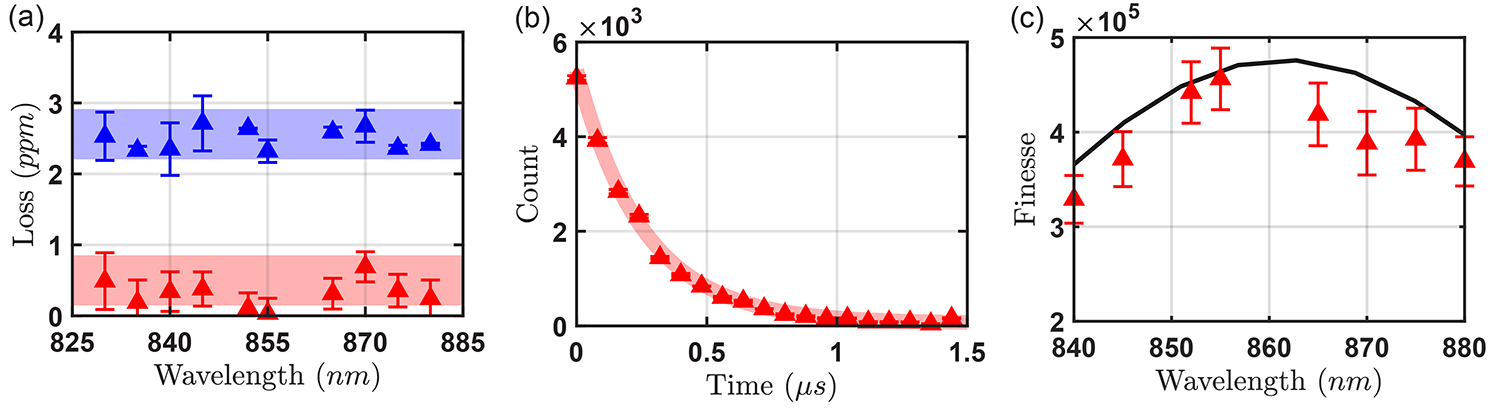}
\caption{Loss partition in cavity ring-down spectroscopy. (a) Absorption-scatter loss before (blue) and after laser annealing (red). At the design wavelength ($852$ nm), we obtain $AS=0.1\pm 0.2$ ppm after laser annealing. (b) Cavity ring-down spectroscopy. We observe the cavity decay rate $\kappa/2\pi=130$ kHz after laser annealing. (c) Ultra-high-finesse $\mathcal{F}$ optical cavity. Together with the results of (a) and transmission loss $T=6.1\pm 0.5$ ppm at $852$ nm, we obtain cavity finesse $\mathcal{F}=(4.4\pm 0.3)\times 10^5$ after laser annealing.}
\label{fig5}
\end{figure*}

As shown in Fig. \ref{fig3}(a), we observe that the oxygen defects are typically found at the interfacial layers between the tantala and silica. To further confirm that these images are the result of the PL from oxygen defect states, we construct the spectrum of the PL via a high-resolution spectrograph with resolution $0.1$ nm, as shown in Fig. \ref{fig4}. In particular, we are able to resolve both mid-gap trap levels at $\sim 567$ nm and  $\sim 647$ nm, consistent to Ref. \cite{Devan2009}. In particular, we observe that the PL signal from the trap level of type $3$ vacancy can extend down to $852$ nm. This implies that optical excitations between the defect levels may lead to optical absorption at the NIR range. Motivated by these measurements and Refs. \cite{Zhang1999,Fleming2000,Gangloff2015}, we perform atmospheric laser-assisted annealing of the tantala with a $10$-mW UV laser at $380$ nm focused to spot size $20\mu m$. In Fig. \ref{fig3}, we demonstrate that out of the $18$ defects, $9$ defects can be recovered with this method. We find that the laser annealing is only effective for oxygen defects on the mirror surface, which hints a potential mechanism of laser-assisted stoichiometry for vacancies exposed to ambient oxygen. Further study is required to examine the physical mechanism behind the laser-assisted annealing.

\section{Ultra-high-finesse optical cavity}\label{UHFcavsection}

By spatially localizing and curing individual defects, we systematically construct an ultra-high-finesse optical cavity with the mode area positioned near defect-free regions. To perform the loss partiion, we follow the protocol of Ref. \cite{Hood2001}. Namely, we detect both the transmitted ($P_t$) and reflected ($P_r$) probe power during a cavity ring-down spectroscopy. Along with the cavity decay rate $\sim \kappa$, this method allows us to obtain
\begin{align}
&\mathcal{F}=\frac{\pi}{\mathcal{L}}= \frac{{FSR}}{2 \kappa},\\
&AS = \frac{2 \pi \kappa}{{FSR}} \left( \frac{P_{\text{r}}-P_{{t}}}{P_{{r}}+P_{\text{t}}} \right), \\
&T = \frac{2 \pi \kappa}{{FSR}} \left( \frac{2 P_{\text{t}}}{P_{{r}}+P_{{t}}} \right),
\end{align}
where the free-spectral range (FSR) is $FSR=c/2L_c$ and the total loss is $\mathcal{L}=AS+T$. In our measurement, we directly probe the $FSR=116.52\pm 0.03$ GHz of the $\sim 1.3$ mm-long cavity with stabilized lasers on a wavemeter. To probe the optical cavity, we use a narrow-linewith ($\sim 100$ Hz) Ti:S laser for the cavity ring-down spectroscopy. The probe field is extinguished within $10$ ns. The two Geiger-mode single-photon APDs are calibrated with intensity stabilized laser, and the uncertainties of the passive losses of the optical setup are accounted as systematic errors in all data. 

Fig. \ref{fig5} presents the loss partitions of the optical cavity for many-body QED experiment \cite{Kong2021}. The test cavity is constructed in a class-$10$ cleanroom environment. For the post-processed mirror samples with the defect free-region near the center, the cavity mode is imaged through the mirror with the resolution $\sim 3\mu m$ limited by the spherical aberrations. The cavity ring-down spectroscopy is performed within a vacuum chamber with moderate pressure $\sim 1$ mTorr to avoid the contributions from the atmospheric losses. In Fig. \ref{fig5}(a), we demonstrate that the post-processing by laser-annealing can make tangible improvement in the absorption-scatter loss of thermally annealed sample. We observe that the absorption-scatter loss at $852$ nm is reduced from $AS=2.6\pm0.4$ ppm to a record-low $AS=0.1\pm 0.2$ ppm, reaching the physical limits of the scattering loss $S\sim 0.14$ ppm of the mirror with the RMS surface roughness $\sim 0.25 \mathring{A}$ determined from atomic force microscopy. The summary of the loss partitions for the original mirrors and the final post-processed mirrors are shown in Table \ref{LTtable}.

\begin{table}[t!]
\begin{center}
    \begin{tabular}[b]{ |c | c | c |c|c|} 
        \hline
        {AS Loss (before)} &{T Loss (before)} &{AS Loss (after)} & {T Loss  (after)} \\
        \hline
        $3.3\pm 0.5$ ppm & $6.1\pm 0.5$  ppm & $0.1\pm 0.2$ ppm & $7.0\pm 0.4$ ppm  \\
        \hline
    \end{tabular}
    \caption{Summary of loss partitions at $852$ nm for ultra-low-loss dielectric mirrors before and after the post-processing (thermal and laser annealing stages). Absorption-scatter (AS) and transmission (T) losses are partitioned with the cavity ring-down spectroscopy based on the method of Ref. \cite{Hood2001}.}
    \label{LTtable} 
    \end{center} 

\end{table}

\section{Conclusion}
In conclusion, we have provided the first microscopic study that connects structural deformation and optical losses of amorphous media. We investigated the key role of oxygen vacancies in tantala for the long-range and short-range structural deformations. At the level of LRO, the dislocation and disorder of amorphous tantala have an indirect quantitative contribution to the absorption losses by the Urbach tails for short-wavelength optical resonators. In the SRO regime, we have investigated the optical transitions between the trap levels of in-plane oxygen vacancies at the individual defect level by photoluminescence spectroscopy and deconvolution microscopy. For long-wavelength ultra-high-finesse optical resonators, optical losses are dominated by direct optical excitations of the defect states. By developing a novel laser-annealing technique, we provide a microscopic tool to recover stoichiometry at the atomistic level. 

Utilizing these understandings, we report the realization of the ultra-high-finesse optical cavity for many-body QED experiments with finesse $\mathcal{F}=(4.4\pm 0.3)\times 10^5$ \cite{Kong2021}, where losses are only limited by the surface qualities over a designated cavity mode volume. Importantly, our measurement of absorption-scatter (AS) loss $AS=0.1\pm 0.2$ ppm improves upon the landmark achievement by Rempe \textit{et al.}, which have held the world record $AS=1.1$ ppm for any optical device over the past three decades \cite{Rempe1992}. Stated differently, we realize a nearly defect-free amorphous optical system, whose optically active atomistic defects could be counted over a region $\sim 1000 \mu m^2$.

\begin{acknowledgments}
This research was supported by the Canada Foundation for Innovation (CFI), the Compute Canada, the IDEX Bordeaux, the Innovation, Science and Economic Development Canada (ISED), the KIST Institutional Program, the Natural Sciences and Engineering Research Council of Canada (NSERC), the National Research Council Canada, the NVIDIA, and the Ontario Ministry of Research and Innovation. The University of Waterloo's  Quantum-Nano Fabrication and Characterization Facility (QNFCF) was used for part of this work. This infrastructure is supported by CFREF-TQT, CFI, ISED, the Ontario Ministry of Research \& Innovation, and Mike \& Ophelia Lazaridis. We acknowledge the CMC Microsystems for the CAD tools and fabrication services. 
\end{acknowledgments}

\end{document}